\documentclass{article}
\begin{document}

\title{Comment on the narrow charmonium state of Belle  at 3871.8 MeV as a deuson}

\author{Nils A. T\"ornqvist\thanks{\tt{e-mail: nils.tornqvist@helsinki.fi}} \\
Department of Physical Sciences, \\ University of Helsinki, POB
64, FIN--00014}

\maketitle

\begin{abstract}
It is pointed out that the  narrow charmonium state at 3871.8 MeV
reported by the Belle collaboration is very likely a $D\bar D^*$
deuteronlike meson-meson state called a deuson. It was predicted
near the $D\bar D^*$ threshold over 10 years ago. Its spin-parity
would be $0^{-+}$ or $1^{++}$ and an important decay mode should
be via $D^0\bar D^{*0}$ to $D^0\bar D^0\pi^0$. Its width to that
channel should then be of the order $50$ keV.

Note: A more detailed update of this unpublished note is in
hep-ph/0402237.

\end{abstract}

No doubt the deuteron is a multiquark state which to a good
approximation can be understood as a proton-neutron system bound
by mainly pion exchange. Many years ago\cite{Tornqvist92}  I asked
the question: for which quantum numbers is the well known pion
exchange mechanism attractive and comparatively large? This was
found to be the case for several light meson-meson channels  with
quantum numbers where problematic resonances have been seen. Then
I predicted\cite{Tornqvist94}  for heavy mesons using a similar
framework as for the deuteron\cite{Glen}  many deuteronlike
states, which are listed  in table 1. See also
Refs.\cite{Mano,Karl}.

\begin{table}[htb]
\begin{tabular}{l l l |l l l }
\hline
       Composite  & $J^{PC}$\ \ \ &     Mass [MeV] &  Composite  &
$J^{PC}$\ \ \ & Mass [MeV]
\\
\hline\hline $D\bar D^*$ & $0^{-+}$& $\approx 3870$ & $B\bar B^*$
& $0^{-+}$& $\approx
10545$            \\
      $D\bar D^*$ & $1^{++}$& $\approx 3870 $ &  $B\bar B^*$ &
$1^{++}$& $\approx
10562$                   \\
\hline $D^*\bar D^*$ & $0^{++}$& $\approx 4015 $ & $B^*\bar B^*$ &
$0^{++}$& $\approx
10582$                    \\
$D^*\bar D^*$ & $0^{-+}$& $\approx 4015 $ & $B^*\bar B^*$ &
$0^{-+}$& $\approx
10590$                   \\
$D^*\bar D^*$ & $1^{+-}$& $\approx 4015 $ &  $B^*\bar B^*$ &
$1^{+-}$& $\approx
10608$                    \\
$D^*\bar D^*$ & $2^{++}$& $\approx 4015 $ &  $B^*\bar B^*$ &
$2^{++}$& $\approx
10602$                 \\
\hline
\end{tabular}
\centering \caption{Predicted masses\protect\cite{Tornqvist94} of
heavy  deuteronlike states called deusons. These are close to the
$D\bar D^*$ and the $D^*\bar D^*$ thresholds, and about 50 MeV
below the $B\bar B^*$ and $B^*\bar B^*$ thresholds. All states
have I=0. The mass values were obtained from (a rather
conservative) one-pion exchange contribution only.}
\label{tab:heavydeusons}
\end{table}

 Very recently the Belle collaboration\cite{Belle0308029} reported a new
narrow charmonium state at $3871.8\pm 0.7$ MeV and with a width
smaller than their resolution (or $\Gamma < 3.4 $ MeV). This is
60-100 MeV above the expected spin 2 $c\bar c $ ($^3D_{c2}$)
state\cite{Eichten80,Buchmuller}. The B-factory produces
abundantly $B{^\pm}$ mesons through $e^+e^-\to \Upsilon(4)\to B^+
B^-$ and they see the new state in the $\pi^+\pi^-J/\psi$
invariant mass distribution of $B^\pm$ decay to $K^\pm
\pi^+\pi^-J/\psi$. They find $34.4\pm 6.5$ events and a
$8.6\sigma$ signal significance   for the observed resonance peak.
%It is shown in Fig.\ref{BelleDeuson}a, which can be compared with
%the well known $\psi'$ (Fig.\ref{BelleDeuson}b) seen in the same
%reaction.

This looks very much like one of  the two first deuteronlike
$D\bar D^*$ states at 3870 MeV predicted in table 1 (from table 8
of Ref.\cite{Tornqvist94}). (The states are of course all
eigenstates of C-parity, or $(|D\bar D^*>\pm |\bar DD^*>)/\sqrt 2$
although we above denote them $D\bar D^*$.) We note in particular:

\begin{itemize}

\item Its spin-parity should  be either $0^{-+}$
or $1^{++}$.   For other quantum numbers pion exchange is
repulsive or so weak that bound states should not be expected.

\item No $D\bar D$ nor $B\bar B$
deusons are expected since the three pseudoscalar coupling (in
this case the $DD\pi$ coupling) vanishes because of parity.

\item If isospin were exact the Belle state as a deuson
would be a pure isosinglet with a mass very
close to the $D\bar D^*$ threshold. For  isovector states pion
exchange is generally one third weaker than for isoscalar states.
Therefore all predicted states were isosinglets. (But see the
comment on isospin breaking below).

\item As a deuteronlike state with small binding energy (for
the deuteron it is 2.22 MeV) the Belle state should be large in
spatial size. It should then have a very narrow width since
annihilation of the loosely bound $D\bar D^*$ state to other
hadrons is expected to be small, although states containing the
$J/\psi$ are  favoured compared to states with only light hadrons
due to the OZI rule.

\item A large part of its width should be given by the
instability of its components or the $D^*$ widths, corrected for
phase space effects. An important decay should be via $D^0\bar
D^{*0}$ to $D^0\bar D^0\pi^0$ since the other charge modes lie
about 2 MeV above the resonance. One can, using isospin and $D^*$
width measurements\cite{Anastassov02}, roughly estimate that that
width should be of the order 50 keV.

\item
The observed peak (at 3871.8$\pm 0.7$ MeV) is almost exactly at
the  $D^0\bar D^{*0}$ threshold (3871.2 MeV), while the
$D^+D^{*-}$ channel, 8.1 MeV higher,  is closed by phase space.
\end{itemize}

{\it Isospin breaking} Since the binding energy of a $D\bar D^*$
deuteronlike state is
 of the same order as the isospin mass
splittings one should expect substantial isospin breaking. For a
pure $I=0$ state one has equal contribution of the two components
in $(|D^ 0\bar D^{*0}>+|D^+ D^{*-}>)/\sqrt 2$, but since $D^ 0\bar
D^{*0}$ is 8.1 MeV lighter than $D^+ D^{*-}$ it should have a
greater weight than $D^+ D^{*-}$. This means there is an $I=1$
component in the state. It is interesting to note that then the
decay of the deuson to $\rho^0 J/\psi\to \pi^+\pi^-J/\psi$ would
not be completely forbidden. There are  indications of this in the
experiment\cite{Belle0308029}. On the other hand the decay chain
$\sigma J/\psi\to \pi^+\pi^-J/\psi$, where $\sigma$ is any
isoscalar object, is forbidden by spin-parity for a
$J^{PC}=0^{-+}$ or $1^{++}$ deuson. Thus one should expect some
$\rho^0 J/\psi\to \pi^+\pi^-J/\psi$, although suppressed by
isospin.

The heavier the constituents the stronger is the binding, since
the kinetic repulsion becomes smaller and is more easily overcome
by the attraction from the potential term. Thus as seen from table
1 the $D\bar D^*$ and $D^*\bar D^*$ systems are barely bound but
for $B\bar B^*$ and $B^*\bar B^*$ the binding energy is $\approx
50$ MeV. (The pion is always too light to be itself a
constituent.)

An uncertainty in the calculation\cite{Tornqvist94} was the $D^*$
coupling to $D\pi$, which was modelled from the $NN\pi$ coupling.
We predicted a $D^{*+}\to D^0\pi^+$ width of 63.3 keV in excellent
agreement with the recent measurement of $65\pm 3$
keV\cite{Anastassov02}. This increases the reliability of that
calculation.

We conclude with a few general comments. For flavour exotic
two-meson systems ($I=2$,  double strange, charm  or bottom), such
as $DD^*$ or $B^*B^*$,
 pion exchange is
always either weakly attractive or repulsive  Calculations do not
support such bound states to exist from pion exchange alone, and
shorter range forces are expected to be repulsive. Should $BB^*$
states exist, however (See \cite{Mano}), they would be quite
narrow since they would be stable against strong decays.

Another recently seen state where pion exchange should be
important is the BES state\cite{BES} seen in $J/\psi\to \gamma
p\bar p$ near the $p\bar p$ threshold. Pion exchange should be
attractive for a pseudoscalar $^1S_0$ state, although a 17 MeV
binding energy for a state with mass 1859 MeV seems a little
difficult to obtain with pion exchange alone.

As to the recent narrow charm-strange resonances seen by the BaBar
and CLEO collaborations at 2317 MeV \cite{BaBarcs}and 2460
MeV\cite{CLEOcs} 30-40 MeV below the $DK$ respectively $D^*K$
threshold  pion exchange is not expected to play a dominant role.
These states (if $J^{PC}=0^{++}$, respectively $1^{++}$) should be
distorted from naive expectations by the strong S-wave cusp at the
above mentioned thresholds. See\cite{AmslerTornqvist,Beveren03}

 A more detailed understanding
with further experimental information on the Belle charmonium
state is important. If the above comments are supported by data it
would open up a completely new spectroscopy. It would also throw
new light on many problematic light resonances in particular the
$\eta(1410)$ and  $\eta(1480)$ resonances just above the $K\bar
K^*$ threshold.

\end{document}